\documentclass{aa}
\usepackage{graphicx}
\usepackage{amssymb}
\usepackage{amsmath}
\usepackage{natbib}
\newcommand{\corot}{{\textsc{CoRoT}}}
\newcommand{\cible}{{HD~49385}}

\newcommand{\ind}[1]{_{\rm #1}}

\def\m2s2{\,m$^{2}$\,s$^{-2}$} %m2.s -2
\def\kms{\,km\,s$^{-1}$}       %km.s -1
\def\vsini{$v\sin i$}          %vsini
\def\logg{$\log g$}

\newcommand{\rotshift}{\delta_{n,\ell,m}}

\newcommand{\vmacro}{$v_{\rm macro}$}
\newcommand{\chisq}{$\chi^2$}
\newcommand{\iet}{\,{\sc i}}
\newcommand{\ito}{\,{\sc ii}}
\newcommand{\vwa}{VWA}
\newcommand{\teff}{$T\ind{eff}$}
\newcommand{\templogg}{\textsc{TEMPLOGG}}

\newcommand\T{\rule{0pt}{2.6ex}}
\newcommand\B{\rule[-1.2ex]{0pt}{0pt}}

\begin{document}
\title{Seismic\thanks{Based on data obtained from the CoRoT (Convection, Rotation and planetary Transits) space mission, 
developed by the French Space agency CNES in collaboration 
with the Science Programs of ESA, Austria, Belgium, Brazil, Germany 
and Spain.}
 and spectroscopic\thanks{Based on data obtained using the T\'elescope Bernard Lyot at
Observatoire du Pic du Midi, CNRS and Universit\'e Paul Sabatier,
France.}
 characterization of the solar-like pulsating CoRoT target HD\,49385}
%\subtitle{}
\titlerunning{Seismic and spectroscopic analysis of solar-like pulsating CoRoT\thanks{The CoRoT space mission, launched on 2006 December 27, was delopped an
d is operated by the CNES with participation of the Science Programs of ESA; ESA's RSSD, Austria, Belgium, Brazil, Germany and Spain.}  target HD\,49385}
\author{
S. Deheuvels\inst{1}
\and H. Bruntt \inst{1}
\and E. Michel \inst{1}
\and C. Barban \inst{1}
\and G. Verner\inst{2}
\and C. R\'egulo \inst{3,4}
\and B. Mosser \inst{1}
\and S. Mathur\inst{5}
\and P. Gaulme \inst{6}
\and R.~A. Garcia \inst{7,8}
\and P. Boumier \inst{6}
\and T. Appourchaux\inst{6}
\and		  R. Samadi\inst{1}
\and		  C. Catala \inst{1}
\and        F. Baudin\inst{6}
\and A. Baglin \inst{1}
\and	  M. Auvergne\inst{1}
\and I.~W. Roxburgh \inst{1,2}
\and	  F. P\'erez Hern\'andez\inst{3,4}
}

   \institute{LESIA, UMR8109, Universit\'e Pierre et Marie Curie, Universit\'e Denis Diderot, Observatoire de Paris, 92195 Meudon Cedex, France\\
              \email{sebastien.deheuvels@obspm.fr}
 \and Astronomy Unit, Queen Mary, University of London Mile End Road, London E1 4NS, UK5
 \and Instituto de Astrof\'isica de Canarias, 38205, La Laguna, Tenerife, Spain
 \and Universidad de La Laguna, 38206 La Laguna, Tenerife, Spain
\and Indian Institute of Astrophysics, Koramangala, Bangalore 560034, India
\and Institut d'Astrophysique Spatiale, UMR8617, Universit\'e Paris XI, B\^atiment 121, 91405 Orsay Cedex, France 
\and Laboratoire AIM, CEA/DSM-CNRS-Universit\'e Paris Diderot; CEA, IRFU, SAp, centre de Saclay, 91191, Gif-sur-Yvette, France
\and GEPI, Observatoire de Paris, CNRS, Universit\'e Paris Diderot; 5 place Jules Janssen, 92190 Meudon, France
}

\offprints{S. Deheuvels\\ \email{sebastien.deheuvels@obspm.fr}
}

\date{Submitted ...}%; accepted March 16, 1997}

\abstract{The star HD~49385 is the first G-type solar-like pulsator observed in the seismology field of the space telescope CoRoT. 
The satellite collected 137 days of high-precision photometric data on this star,
confirming that it presents solar-like oscillations. HD~49385 was also observed in spectroscopy with the NARVAL spectrograph in January 2009.}
{Our goal is to characterize HD~49385 using both spectroscopic and seismic data.}
{The fundamental stellar parameters of HD~49385 are derived with the semi-automatic software VWA, and the projected rotational velocity is estimated 
by fitting synthetic profiles to isolated lines in the observed spectrum. A maximum likelihood estimation is used to determine the parameters of the observed p modes.
We perform a global fit, in which modes are fitted simultaneously over nine radial orders, with degrees ranging from $\ell=0$ to $\ell=3$ (36 individual modes).}
{Precise estimates of the atmospheric parameters ($T\ind{eff}$, [M/H], $\log g$) and of the \vsini\ of \cible\ are obtained. The seismic analysis of the
star leads to a clear identification of the modes for degrees $\ell=0,1,2$. Around the maximum of the signal ($\nu\simeq1013\,\mu$Hz), some peaks are found significant and compatible with the expected
characteristics of $\ell=3$ modes. Our fit yields robust estimates of the frequencies, linewidths and amplitudes of the modes. 
We find amplitudes of $\sim5.6\pm0.8$ ppm for radial modes at the maximum of the signal. The lifetimes of the modes range from one day (at high frequency) to a bit more than two days
(at low frequency). Significant peaks are found outside the identified ridges and are fitted. They are attributed to mixed modes.}
{}


\maketitle
%________________________________________________________________
\section{Introduction}

In the Sun, oscillations are excited by the turbulent motions in the outer part of the external convective envelope and are further propagated into the interior of the star.
The study of these oscillations has yielded constraints on the inner structure of the Sun, allowing us to estimate
the sound speed and density profiles (\citealt{2003ApJ...591..432B}, \citealt{2001ApJ...555L..69T}), the position of the base of the convective zone (\citealt{1991ApJ...378..413C}), and
the rotation profile (\citealt{2003ARA&A..41..599T}, \citealt{2008A&A...484..517M}). However, the very low amplitude of these oscillations (a few ppm in photometry) makes it very challenging to
detect and analyze them in other stars than the Sun.

Achieving a better understanding of the interiors of solar-like pulsations is one of the main objectives of the 
space mission \corot\ (\textbf{Co}nvection, \textbf{Ro}tation and planetary \textbf{T}ransits).
\corot\ is a space telescope performing  high-precision photometry over quasi-uninterrupted long observing runs (\citealt{Baglin06}).
Solar-like oscillations have already been studied in several other stars with \corot\ data. 
The analyses of these stars have encountered difficulties identifying the degrees of the modes, either because of a too low signal-to-noise
ratio (HD~175726: \citealt{2009A&A...506...33M}, HD~181906: \citealt{2009A&A...506...41G}), or because of a too short lifetime of the modes, 
inducing large mode linewidths (HD~49933: \citealt{2008A&A...488..705A}, HD~181420: \citealt{HD181420} ).

The star HD 49385 is the first G-type solar-like pulsator observed in the seismology field of \corot. It is cooler than the solar-like pulsators previously analysed with \corot\ data, and probably
more evolved (at the end of the Main Sequence or shortly after it). The choice of \cible\ as a \corot\ target has motivated us
to lead spectroscopic observations, performed with the NARVAL spectrograph at the Pic du Midi Observatory. The fundamental parameters of \cible\ are derived
from these observations, as described in Sect. \ref{sect_param}. The photometric observations with \corot\ are presented in Sect. \ref{sect_obs}.
Section \ref{sect_rot} presents the study of the low-frequency part of the power spectrum, in search of a signature of the stellar rotation. The extraction of p-mode
parameters is described in Sect. \ref{sect_analysis}, and Sect. \ref{sect_concl} is dedicated to conclusions.

%__________________________________________________________________
\section{Stellar parameters of \cible \label{sect_param}}

The solar-like pulsator \cible\ is a G0-type star with an apparent magnitude of $m\ind{V}=7.39$ (\textit{uvby} catalog, \citealp{1998A&AS..129..431H}).
To determine the fundamental parameters of HD~49385 we analysed a high-quality spectrum from the NARVAL spectrograph mounted on the 2-m~Bernard Lyot Telescope
at the Pic~du~Midi Observatory. We stacked two spectra collected on 2009 January 10, totalizing an exposure time of 6\,600 s.
The spectrum was normalized by identifying continuum windows in a synthetic spectrum and fitting a  
low-order spline through these points. This was done order-by-order and we made sure the line depths agreed for the overlapping part of
adjacent echelle orders. We measured a mean signal-to-noise (S/N) ratio of $600$ in the continuum for several line-free regions
in the range $5\,000$--$7\,000$\,\AA\  for a data sampling of $2.7$ data points per resolution element ($R=65\,000$).

\subsection{Temperature, $\log g$ and metalicity \label{sect_teff}}

%A paragraph written by Hans Bruntt will detail the work done to determine the following parameters:
%$T\ind{eff} = 6095\,\pm\,50$ K \\
%$\log g = 4.0 \, \pm \, 0.08$  \\
%$[$Fe/H$]  = 0.08 \, \pm \, 0.06$ dex  \\
%$v\sin i = 3.5 \, \pm \, 0.8 $ \kms  \\

We used the semi-automatic \vwa\ software (\citealt{bruntt09}) to fit synthetic profiles for more than 600 lines in the range from $4\,135$ to  
$8\,545$\,\AA. We used a differential approach, meaning all abundances are measured relative to exactly the same lines in the solar spectral atlas from
\cite{1984sfat.book.....K}. This differential approach is described in more detail by \cite{bruntt08}. As part of the analysis, we determined the atmospheric model
parameters \teff, \logg, and the microturbulence. This was done by requiring that abundances determined from Fe\iet\ lines do not correlate with the equivalent
width (EW) or the excitation potential (EP). Furthermore we required that the same mean abundances are measured from the Fe\iet\ and Fe\ito\ lines ("ionization balance").
We estimated the uncertainty on the model parameters by adjusting them until the correlations of Fe\iet\ with EW or EP became significant or the ionization
balance deviated (see \citealt{bruntt08} for details). We stress that the uncertainties are strictly internal errors because the underlying assumption is that the model
atmosphere represents the star. Any systematic error in the temperature profile or departures from local thermal equilibrium (LTE) will likely affect the results.
We therefore quadratically added 50~K and 0.05 dex to the uncertainty on \teff\ and \logg. The spectroscopic parameters are listed in Table~\ref{tab_fund}.

As an additional check of the surface gravity we fitted the wings of the pressure sensitive Ca lines at $6122.2$ and $6162.2$\,\AA. This was
done by first adjusting the van der Waals constants to fit the solar spectrum for the canonical value $\log g=4.438$. The fitted values for the two lines
for HD~49385 are $\log g = 3.96\pm0.12$ and $4.05 \pm 0.06$, which agrees very well with the Fe\iet/Fe\ito\ ionization balance.
The low value we obtain for the $\log g$ of \cible\ indicates that it certainly is an evolved object.

\begin{table}
  \centering
  \caption{Spectroscopic parameters of \cible.
  \label{tab_fund}}
\begin{tabular}{ c  c }
\hline \hline
\T $T\ind{eff}$ & $ 6095 \pm65 $\,K  \\
$\log g   $   & $ 4.00 \pm 0.06$ \\
${\rm [M/H]}$ & $+0.09 \pm 0.05$ dex \\
\B \vsini & $2.9_{-1.5}^{+1.0}$\,\kms \\
\hline
\end{tabular}
\end{table}

The parameters agree well with the calibration of the Str\"omgren indices. From \templogg\ (\citealt{rogers95}, \citealt{2001fcm..book...39K}) we get $T_{\rm eff} =  6345 \pm 150$\,K,
$\log g =  4.22 \pm 0.15$, and $[{\rm Fe/H}] = +0.21 \pm 0.10$. \templogg\ determines a significant interstellar reddening $E(b-y) = 0.020\pm0.007$.
If we assume zero interstellar reddening, we get from the $V-K$, using the \cite{2005ApJ...626..465R} calibration, $T_{\rm eff} = 5958 \pm 64$\,K and $6035\pm92$\,K from $b-y$.
The uncertainties include the calibration uncertainty and the photometric uncertainty on the indices. Finally, \cite{2004A&A...427..933K} used line-depth ratios
to determine effective temperatures of stars and found $T_{\rm eff} = 6052 \pm 6.7$\,K (internal error). They also determined \logg\ and $[$Fe/H$]$, but did not
give uncertainties on these two parameters: $\log g =  4.00$ and $[{\rm Fe/H}] = +0.10$. In summary, our result from \vwa\ agrees well with the
photometric calibrations and the line-depth ratio method for \teff.

We found a high abundance of lithium in HD~49385 of $+1.08\pm0.10$
(relative to the Sun). This value appears to be typical for a dwarf
star with solar metalicity (e.g., \citealt{boesgaard05}). We used the
line list from \cite{ghezzi09}, but did not include the relatively weak
molecular CN bands;  hence the Li abundance may be slightly
overestimated. The Li feature around 6707.8 is known to be strongly affected by
NLTE effects, but when using 3D instead of 1D atmosphere models these
two effects appear to cancel to first order (\citealt{asplund05}). In other
words, because we are using 1D LTE models,
the imposed error should be relatively small. Furthermore, because we
are calculating abundances relative to the same Li line feature as in the
Sun, our estimate of [Li] should be robust as the atmospheric parameters
of the two stars are quite similar.

\begin{figure}
\centering
\includegraphics[width=3.4cm,angle=90]{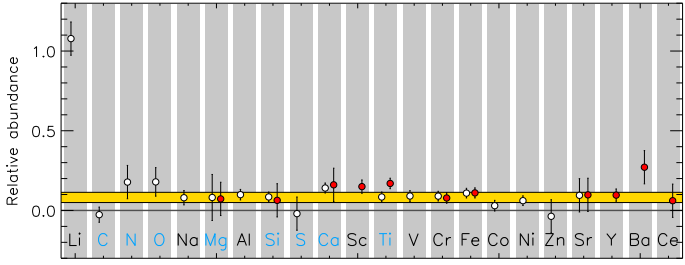}
\caption{Relative abundances of 22 elements measured in \cible.
Open and solid circles are mean values for neutral and
singly ionized lines, respectively.
The yellow horizontal bar marks the mean metalicity with 1-$\sigma$ uncertainty range.
\label{fig_vwa}}
\end{figure}

The abundance pattern we determine is shown in Fig.~\ref{fig_vwa} and listed in Table~\ref{tab_ab}.
The overall metalicity is determined as the mean
of the metals with at least 10 lines used in the analysis
(Si, Ca, Ti, V, Cr, Fe, Co, Ni) giving ${\rm [M/H]} = 0.09\pm0.05$.
This range is marked by the horizontal bar in Fig.~\ref{fig_vwa}.

\subsection{Projected rotational velocity \label{sect_vsini}}

Because we have a relatively high-resolution spectrum with very high S/N we were able to estimate the projected rotational velocity (\vsini).
We fitted synthetic profiles to ten isolated lines in the range $4\,600$--$6\,800$\,\AA. We calculated the synthetic profiles with the SYNTH software \cite{1996A&AS..118..595V},
thus taking into account weak blends. We considered three broadening mechanisms which were convolved with the synthetic profiles: (1) the instrumental resolution,
(2) macroturbulence (\vmacro) and (3) \vsini. The instrumental power of resolution of NARVAL is $R=65\,000$, but to check this value we fitted six narrow telluric lines around 6290\,\AA. We found that
a Gaussian function with a FWHM of 6290\,{\rm \AA}/65\,000 =  0.097\,\AA\ fitted these lines very well and we used the same value of $R$ for the stellar lines.
For \vsini\ and \vmacro\ we calculated a $17 \times 17$ grid of convolved profiles for 0--8\,\kms\ in steps of 0.5\,\kms. To identify the best combination of \vsini\ and \vmacro\ we compared
the synthetic profile and the observed spectrum by computing the \chisq\ value.

\begin{figure}
\centering
\includegraphics[width=8.5cm]{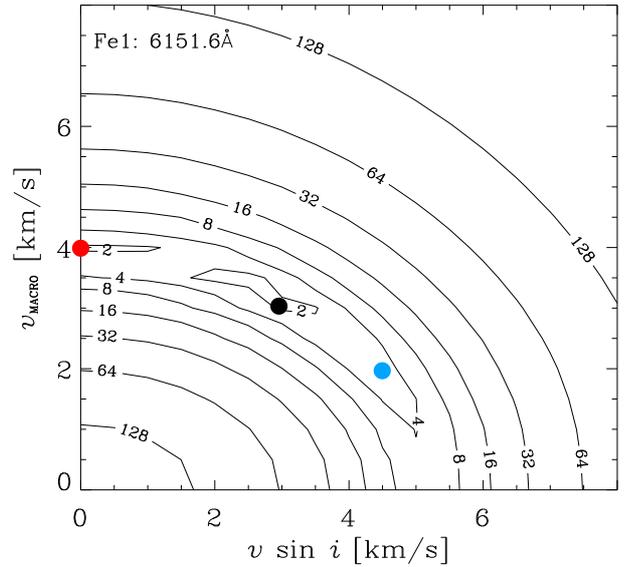}
\caption{Example of the $\chi^2$ contour surface
used to determine $v \sin i$ and $v_{\rm macro}$ for
the {Fe \sc i} line $\lambda6151.6$\,\AA.
The three circles mark the minimum of the surface and
the 1-$\sigma$ uncertainty range. The corresponding
three synthetic spectra are compared to the observed
line in Fig.~\ref{fig_synth}.
\label{fig_contour}}
\end{figure}

\begin{figure}
\centering
\includegraphics[width=8.5cm]{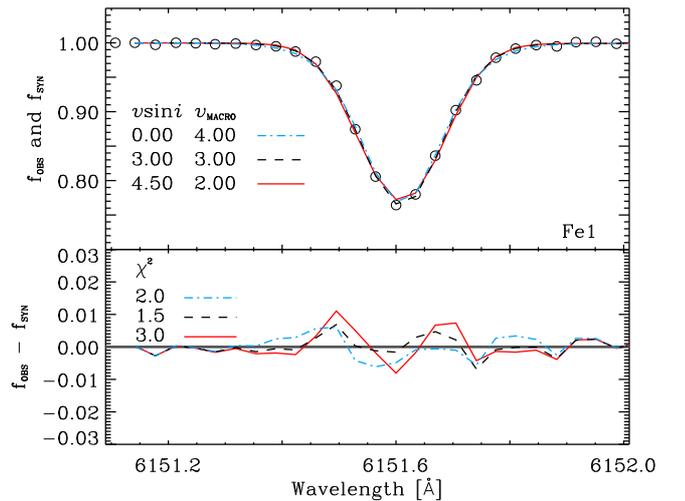}
\caption{The upper panel shows the fit of three synthetic profiles
to the observed line (marked by open circles). Each profile was
convolved with different combinations of $v \sin i$ and $v_{\rm macro}$
as given in the top panel (cf. Fig.~\ref{fig_contour}). The lower panel shows the
difference between the observed and synthetic profiles,
indicating a slightly better fit for the profile represented by a dashed line
(the $\chi^2$ value for each fit is given in the lower panel).
\label{fig_synth}}
\end{figure}

In Fig.~\ref{fig_contour} we show an example of the \chisq\ contour for the Fe\iet\ line at 6156.1\,\AA. The plots for all ten lines are very similar and show the same
strong correlation between \vsini\ and \vmacro. We marked three points on the contour in Fig.~\ref{fig_contour} with filled circles, and the synthetic profiles corresponding to these
grid points are shown in Fig.~\ref{fig_synth}. The observed spectrum is shown with open circles. From our analysis we can put a firm upper limit on \vsini\ of 5\,\kms,
corresponding to $v_{\rm macro}=0$\,\kms. The values that provide the best fit based on the analysis of the \chisq\ contours for all ten lines are $v \sin i      = 2.9_{-1.5}^{+1.0}$\,\kms\ and
$v_{\rm macro} = 3.4^{+0.3}_{-0.7}$\,\kms.

\begin{table}
  \centering
  \caption{Abundances of 22 elements in HD~49385 relative to the Sun.
The third column lists the number of lines used to
determine the mean abundance listed in the second column.
The quoted uncertainties are intrinsic RMS errors on the mean value.
  \label{tab_ab}}
 % \setlength{\tabcolsep}{3pt} % narrow table: default is tabcolsep = 6pt
 % \begin{footnotesize}
\begin{tabular}{l l r    |    l l r}
\hline \hline
\T \B    El.        & $\Delta A$ & $n$ & El.        & $\Delta A$ & $n$ \\
\hline
\T   {C  \sc   i} &     $ -0.03\pm0.05  $ &   9  &  {V  \sc   i} &     $ +0.09\pm0.03  $ &  12  \\
   {N  \sc   i} &     $ +0.18        $  &   1  &    {Cr \sc   i} &     $ +0.09\pm0.03  $ &  18  \\
   {O  \sc   i} &     $ +0.18\pm0.17 $  &   4  &     {Cr \sc  ii} &     $ +0.08\pm0.03  $ &   7  \\
   {Na \sc   i} &     $ +0.08\pm0.04  $ &   4  &     {Fe \sc   i} &     $ +0.11\pm0.03  $ & 336  \\
   {Mg \sc   i} &     $ +0.08\pm0.14  $ &   4  &     {Fe \sc  ii} &     $ +0.11\pm0.03  $ &  27  \\
   {Mg \sc  ii} &     $ +0.07\pm0.03  $ &   2  &     {Co \sc   i} &     $ +0.03\pm0.03  $ &  12  \\
   {Al \sc   i} &     $ +0.10\pm0.03  $ &   4  &    {Ni \sc   i} &     $ +0.06\pm0.03  $ &  81  \\
   {Si \sc   i} &     $ +0.08\pm0.03  $ &  38  &     {Zn \sc   i} &     $ -0.04       $   &   2  \\
   {Si \sc  ii} &     $ +0.06\pm0.04  $ &   2  &    {Sr \sc   i} &     $ +0.09        $  &   1  \\
   {S  \sc   i} &     $ -0.02       $   &   2  &     {Sr \sc  ii} &     $ +0.10         $ &   1  \\
   {Ca \sc   i} &     $ +0.14\pm0.03 $  &  13  &     {Y  \sc  ii} &     $ +0.10\pm0.04 $  &   7  \\
   {Ca \sc  ii} &     $ +0.16\pm0.03  $ &   2  &     {Ba \sc  ii} &     $ +0.27        $  &   1  \\
   {Sc \sc  ii} &     $ +0.15\pm0.04  $ &   6  &     {Ce \sc  ii} &     $ +0.06         $ &   1  \\
   {Ti \sc   i} &     $ +0.08\pm0.03  $ &  30  &   {Li \sc   i} &  $+1.08$   &   1  \\
 \B  {Ti \sc  ii} &     $ +0.17\pm0.03  $ &   9  &  &  &  \\
\hline   
\end{tabular}
%\end{footnotesize}
\end{table}

\subsection{Luminosity}

Using the Hipparcos parallax of the star, $\pi=13.91\,\pm\,0.76$ mas
(\citealt{2007A&A...474..653V}), we could derive the absolute visual magnitude of the object $M\ind{V}=3.11\pm0.12$. The bolometric correction was determined by interpolating in the grid 
provided by \cite{1998A&A...333..231B} for appropriate values of temperature and $\log g$ for \cible. We obtained $BC\ind{V}=-0.029\,\pm\,0.006$, which yields a luminosity $\log(L/L_{\odot})=0.67\,\pm\,0.05$. 
The values of $T\ind{eff}$ and $L$ provided an estimate of the radius $R/R_{\odot}=1.94\pm0.15$.

\section{\corot\ observations \label{sect_obs}}

\begin{figure}
\centering
\includegraphics[width=8.5cm]{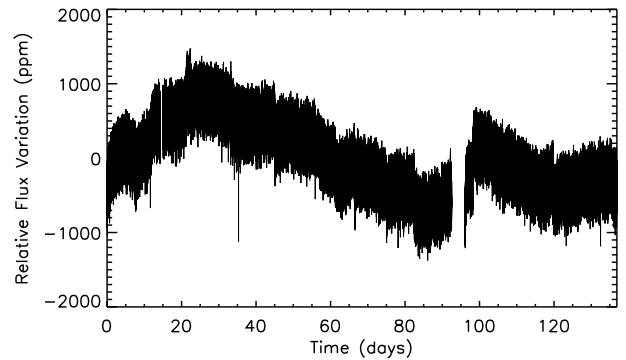}
\caption{Lightcurve of 137 days of \corot\ observations on \cible, detrended using instrumental information.
\label{fig_lightcurve}}
\end{figure}

The star \cible\ was one of the targets of the second Long Run in the asteroseismology field of the space mission \corot. 
136.9 days of photometric data have been collected from October 2007 to March 2008,
with a duty cycle of 88.2\%. After being corrected from known instrumental effects as described in \cite{2007astro.ph..3354S},
the lightcurve consists of a series of equally-spaced measurements (32 s) in the heliocentric frame (also known as N2 data).
Small gaps remain in the lightcurve (a few minutes long) mainly due to the passage of the \corot\ satellite in the South Atlantic Anomaly (SAA).
A larger gap (3.5 days) was caused by an unexpected reset of the DPU (Digital Processing Unit), which was probably the consequence of the impact of an energetic particle, because this event
occured while the satellite was crossing the SAA. These gaps were linearly interpolated, based on the measurements collected 1\,000 s on either sides of the gap. 
A low-frequency trend due to the aging of the CCD was also corrected by removing a linear gain variation of $-5.32\,10^{-5}$ day$^{-1}$ (see \citealt{2009A&A...506..411A}
for more details). The resulting lightcurve is shown in Fig. \ref{fig_lightcurve}.

\begin{figure}
\centering
\includegraphics[width=8.5cm]{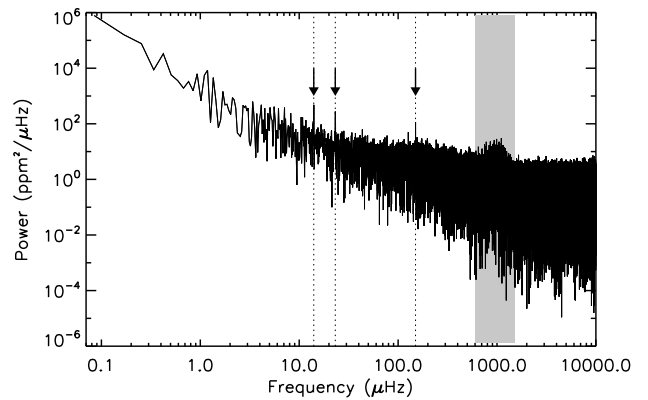}
\caption{Power spectrum of 137 days of \corot\ data for \cible. The peaks indicated by arrows are from left to right: the 14-$\mu$Hz-peak discussed in Sect. \ref{sect_rot},
a peak at $2\nu\ind{day}=23.1\,\mu$Hz, and a peak at $\nu\ind{orb}-\nu\ind{day}=150.1\,\mu$Hz. The grey area shows the excess of power due to the p-mode oscillations.
\label{fig_speclog}}
\end{figure}

Figure \ref{fig_speclog} shows the power spectrum of the lightcurve, computed with the Fast Fourier Transform algorithm (FFT). It was normalized so that the integrated power spectrum
from 0 to twice the Nyquist frequency corresponds to the variance of the time series. At low frequency the background rises due to stellar granulation and possibly
to stellar activity (see discussion in Sect. \ref{sect_rot}). The signature of perturbations due to the orbit of the satellite ($\nu\ind{orb}\simeq161.7\,\mu$Hz) and the day ($\nu\ind{day}\simeq11.6\,\mu$Hz) 
remains in the power spectrum. These perturbations were expected (see \citealp{2009A&A...506..411A}) and are of relatively low amplitudes (a few tens of ppm$^2/\mu$Hz for the
strongest ones at low frequency). They appear as combinations of the harmonics of both frequencies. For instance, a peak appears at 150.1 $\mu$Hz (corresponding to $\nu\ind{orb}
-\nu\ind{day}$). Finally, a broad excess of power, centered around 1 mHz,  is clearly seen in the power spectrum and corresponds to the acoustic modes of \cible, which are studied in Sect. \ref{sect_analysis}.

\section{Search for a signature of stellar rotation in the Fourier domain \label{sect_rot}}

The measurement  of the rotation period from the lightcurve analysis was
found to be useful in previous analyses of p-mode parameters
 (see \citealt{2008A&A...488..705A}, \citealt{HD181420}, \citealt{2009A&A...506...33M}).
It indeed provides a direct measurement of the surface rotation
period and hence a first estimate of the rotational splitting value.

 Contrary to other solar-like pulsating \corot\ targets, no immediate signature of the activity can be seen in the lightcurve
(see Fig. \ref{fig_lightcurve}). Only a long period trend stands out. The same trend was also found in several stars
observed on the same CCD during the same specific observation run. Its instrumental or environmental origin is thus established
and is currently under study.

%, which cannot be related to rotation. Indeed, the estimates we
%obtained for the radius and the projected velocity of \cible\ give us an upper limit of 55 days for the rotation period of the star,
%through the relation $T= 2\pi R \sin i/(v\sin i)$.
%\textbf{A long period unrelated to stellar rotation was also observed in the \corot\ lightcurve of HD~181420 (\citealt{HD181420}). 
%At this stage, it is difficult to say whether these periods are stellar, environmental or instrumental effects.}
%We then search for the signature of the star rotation in the low frequency part of the power spectrum. 

We first identified a peak around $0.4\,\mu$Hz (period of about 29 days), the significance of which is hard to assess 
because it presents a relatively low amplitude in a frequency domain
where the intrumental noise can dominate the power spectrum.
We performed a time-frequency analysis with Morlet wavelets (as described in \citealt{morlet}) which showed that this peak presents a certain stability in time. But because the periods we were
testing correspond to a non-negligible fraction of the total observation period, the cone of influence (\citealt{1998BAMS...79...61T}) prevented us from drawing any conclusion.

A prominent peak can also be detected at a frequency of $14\,\mu$Hz (period of about 0.8 day) with a signal-to-noise ratio of more than 10.
This peak is however unlikely to be directly related to rotation. Indeed, let us assume \textit{ad absurdum} that the rotation frequency is $14\,\mu$Hz. 
Our estimate of the radius implies a surface velocity of more than 120 \kms. To obtain the measured $v\sin i$,
we would need a value of the inclination angle lower than $2^{\circ}$, which means that the star is seen pole-on. In this case
it is extremely difficult to derive the signature of the activity from the light curve (\citealt{Mosser09}) and no peak related to the rotation period can be
detected in the power spectrum, which contradicts our assumption.
In the power spectrum of HD~49933, which is a target of the seismo-field for the same run, we also found a small peak around $14\,\mu$Hz,
suggesting this peak might be instrumental. However, the frequencies do not match exactly, and the peak observed for HD~49933 does not appear to
be significant. The origin of this peak remains unclear. But it cannot be related to rotation.

%, which could explain
%the small amplitude of the peak at $14\,\mu$Hz. It is also difficult to
%disentangle the rotation period from the spot lifetime, so that
%identifying the peak to the exact signature rotation is not possible.

In summary, we found no clear signature of the stellar rotation of \cible. The low-frequency part of the power spectrum can therefore not be used 
to obtain a first estimate of the rotational splitting. We may however infer from the absence of a clear activity signal either that
the star has a low activity, which cannot be detected even with the
\corot\ sensitivity, or that the inclination of the star is close to pole-on.

\section{Analysis of p mode oscillations \label{sect_analysis}}

\begin{figure}
\centering
\includegraphics[width=8.5cm]{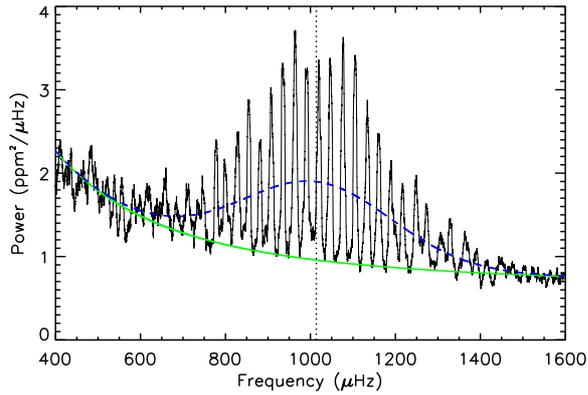}
\caption{Smoothed power spectrum of \cible\ (with a 10 $\mu$Hz boxcar) in the frequency range of the oscillations. The green curve represents the fitted background (granulation and
white noise components). The blue dashed curve corresponds to a Gaussian fit of the raw power spectrum giving $\nu\ind{max}=1013\,\mu$Hz (dotted line). 
\label{fig_signal}}
\end{figure}

Figure \ref{fig_signal} shows the power spectrum smoothed with a 10-$\mu$Hz boxcar. The comb-like
structure of this excess of power is typical of p-mode oscillations. This section presents the analysis of these modes.

\subsection{Frequency of maximum power and large spacing  \label{sect_ls}}

To estimate the frequency of the maximum of signal, we fitted a Gaussian profile on the raw power spectrum 
in the frequency range of the oscillations and above a fitted background (fit described in Sect. \ref{sect_fit}). 
The result is shown in Fig. \ref{fig_signal}. We obtained as the frequency of maximum power $\nu\ind{max}=1\,013\pm3\,\mu$Hz.

In Fig. \ref{fig_signal} a comb-like structure is clearly visible between 600 and $1400\,\mu$Hz.
The autocorrelation of the power spectrum provides a first estimate of the mean value of the large spacing $\overline{\Delta\nu}\simeq56\,\mu$Hz.
%To determine a mean value of the large spacing, we searched for equidistances in the power spectrum by computing its autocorrelation function on this frequency range (see Fig. \ref{fig_autocor}).
%The third peak corresponds to the large spacing $\Delta\nu=\nu_{n,\ell}-\nu_{n-1,\ell}$. 
%A peak with a flat top appears at about 26 to 30 $\mu$Hz. This flat top is the consequence of the merging of two different peaks, corresponding to two quantities close to half the large separation:
%$D_{01}=\nu_{n,0}-\nu_{n-1,1}$ and $D_{10}=\nu_{n,1}-\nu_{n,0}$. 
%From the third peak, we obtain a first estimate of the 
%mean value of the large spacing to be about $56\,\mu$Hz.

%\begin{figure}
%\centering
%\includegraphics[width=8.5cm]{fig_autocor.ps}
%\caption{Autocorrelation of the power spectrum over the frequency domain $[600,1\,400]\,\mu$Hz (grey curve). A smooth of the autocorrelation function has been overplotted (black curve).
%The dotted line indicates the inferred mean value of the large spacing $\overline{\Delta\nu}=56\,\mu$Hz.
%\label{fig_autocor}}
%\end{figure}

\subsection{ Mode identification \label{sect_id}}

\begin{figure}
\centering
\includegraphics[width=8.5cm]{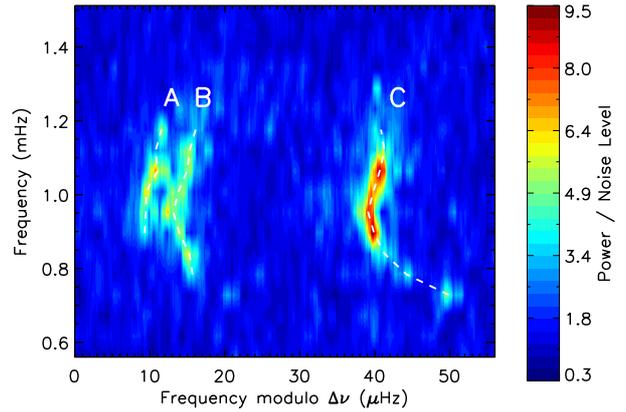}
\caption{\'Echelle diagram of \cible\ computed with a large spacing of $\Delta\nu=56\,\mu$Hz. The power spectrum is smoothed with a 1-$\mu$Hz boxcar and
normalized by the background (estimated in Sect. \ref{sect_method}). Three clear ridges appear, labeled here as ridges A, B, and C. The white dashed lines
correspond to the frequencies obtained from a fit to the data presented in Sect. \ref{sect_fit}, and summarized in Tables \ref{tab_param0A} and \ref{tab_param12A}.
\label{fig_echelle}}
\end{figure}

We used the value of the large spacing given in the previous section to build the \'echelle diagram shown in Fig. \ref{fig_echelle}. 
It is obtained by piling onto one another sections of 56 $\mu$Hz of the power spectrum.
Three clear ridges appear in the \'echelle diagram: two neighboring ones, referred to as ridges A and B in Fig. \ref{fig_echelle}, and a third one (ridge C).

The pattern we observe in Fig. \ref{fig_echelle} is the one we expect to observe for low-degree high-radial-order modes in solar-like pulsators: two main ridges separated by about half a large
spacing corresponding to $\ell=0$ and $\ell=1$ modes, and alongside the $\ell=0$ ridge a fainter one corresponding to $\ell=2$ modes.
It then seems reasonable to identify the ridges as follows: neighboring ridges A and B correspond to $\ell=2$ and $\ell=0$ modes, respectively, and ridge C
to $\ell=1$ modes. We note that \cite{2009A&A...508..877M} and \cite{bedding10} obtained the same identification for the degrees of the ridges using different approaches.
The following analysis is based on this identification. However, since the previous analyses of \corot\ solar-like targets encountered
ambiguity indentifying the modes (see \citealt{HD181420}, \citealt{2009A&A...506...41G}), we decided to also consider the possibility of the alternate identification 
(for which ridge C corresponds to $\ell=0$ modes), which is discussed in Sect. \ref{sect_alternate}. \\

\begin{figure}
\centering
\includegraphics[width=8.5cm]{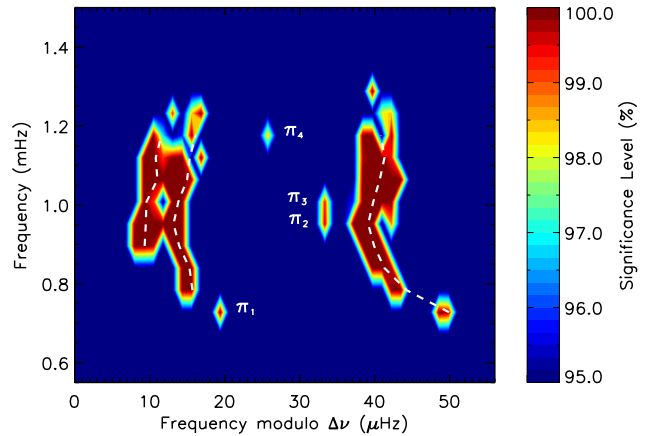}
\caption{\'Echelle diagram of the peaks with a significance greater than 95\% in the power spectrum smoothed over $n=15$ bins ($\sim1\,\mu$Hz). Four
peaks appear, which do not belong to the identified ridges A, B, or C, and are labeled as $\pi_{1,2,3,4}$.
\label{fig_schusters}}
\end{figure}

Figure \ref{fig_echelle} also shows fainter peaks outside the identified ridges. We here study their significance by testing the H$_0$ hypothesis
in the frequency range of the oscillations.
Because we searched for short-lived modes, we binned the power spectrum over $n$ bins, as prescribed in \cite{app04}. The size of the boxcar ($n=15$ bins $\sim 1\,\mu$Hz) 
for the binning was chosen to approximately match the expected linewidth for the p modes in solar-like pulsators. 
In this case, the noise is distributed as a $\chi^2$ with $2n$ degrees of freedom.
We then applied the H$_0$ hypothesis, \textit{i.e.} we supposed that the observations are due to pure noise.
For each peak we computed the probability $p$ that pure noise is able to reach the height of the peak over the whole studied window.
Low values of $p$ indicate a poor compatibility between the observations and noise, and the significance of the peaks is defined as $1-p$.
We show in Fig. \ref{fig_schusters} all the peaks with a significance higher than 95\%. 

We notice four peaks which appear with a significance above 98\%, and lie outside the identified ridges. 
They are labeled as $\pi_1$, $\pi_2$, $\pi_3$ and $\pi_4$ on Fig. \ref{fig_schusters}.
The peak $\pi_4$, clearly outside the pattern expected from asymptotic theory, could correspond to a mixed mode, given that \cible\ is an evolved object.
The peak $\pi_1$ might be an $\ell=0$ mode, assuming that the curvature of ridge B changes at low frequency. 
This hypothesis seems to be corroborated by the fact that such a change exists in ridge C.
However, a more precise inspection shows that including $\pi_1$ in ridge B generates an abrupt step in the $\ell=0$ large separation, whereas the 
$\ell=1$ large separation varies quite smoothly. The peak $\pi_1$ could also be an $\ell=1$ mixed mode in avoided crossing.
Indeed, at low frequency, the $\ell=0$ and $\ell=1$ ridges have different curvatures. This can
be seen in Fig. \ref{fig_echelle} and will be confirmed by the large separation profiles derived from the fitted mode frequencies (see 
Sect. \ref{sect_mode_param}). It was shown in \cite{deheuvels10} that a low-degree $\ell$ avoided crossing creates a characteristic
distortion in the ridge of degree $\ell$. If we assume that $\pi_1$ is an $\ell=1$ mixed mode in avoided crossing, the expected curvature of the $\ell=1$ ridge is very similar to 
that of the observed ridge. We cannot establish this identification more firmly at this stage though. Because of the uncertainty regarding the identification 
of the $\pi_1$ mode, we preferred to consider and fit it individually (see Sect. \ref{sect_fit}).
The peaks $\pi_2$ and $\pi_3$ are located at about 7 $\mu$Hz on the left of ridge C.
We studied the possibility for these peaks to correspond to mixed modes. 
In this case, they would have to be of different degree $\ell$. Indeed, g modes of same degree and 
increasing radial order have frequencies increasingly close to each other. Two mixed 
modes of same degree $\ell$ spaced by $\Delta\nu$ would imply many more mixed modes in the 
ridge of degree $\ell$ at lower frequency. It is clear from ridges $\ell=1$ and $\ell=2$ that this is not 
the case. And if $\pi_2$ and $\pi_3$ were mixed modes of different degree, 
we would expect them to have different amplitudes, which does not seem to be the case.
Their position in the \'echelle diagram on the left of ridge C suggests that they could also be $\ell=3$ modes.
Several reasons led us to favor this identification. First, these two modes appear vertically aligned in the \'echelle diagram
and  around the maximum of the oscillations,
\textit{i.e.} at the place where it is most likely to observe $\ell=3$ modes if they are present.
Then, $\pi_2$ and $\pi_3$ might be part of a fainter ridge (referred to as ridge D), because other peaks, which have a lower significance but seem to follow the 
alignment, appear in Fig. \ref{fig_echelle}. Finally, the distance between ridges C and D is also consistent
with this identification. Indeed, the asymptotic theory gives an expression for the average distance between the ridges
\begin{equation}
\langle \delta\nu_{n,\ell} \rangle \equiv  \nu_{n,\ell}-\nu_{n-1,\ell+2}  = (4\ell+6) D_0,
\end{equation}
where the expression of $D_0$ is found in \cite{coursCD}. By definition, the quantity $\langle \delta\nu_{n,\ell=0} \rangle$ 
corresponds to the distance between the $\ell=0$ and $\ell=2$ ridges, and
$\langle \delta\nu_{n,\ell=1} \rangle $ corresponds to the distance between the $\ell=1$ and $\ell=3$ ridges. Figure \ref{fig_echelle} 
shows that we have $\langle \delta\nu_{n,\ell=0}  \rangle \simeq5\,\mu$Hz.
We therefore expect $\ell=3$ modes to be approximately 8 $\mu$Hz on the left of $\ell=1$ modes, which approximately matches the position of ridge D.

The significance is an efficient tool to select peaks, but
it is not a guarantee for the detection of a signal, as stressed in \cite{app09}. To confirm the existence of a signal, one needs to
compute the posterior probability $P(H_0|X=x)$. The details of the calculation along with the
assumptions we make are given in Appendix \ref{app_bayes}. 
For $\pi_1$ and $\pi_4$, assumed to be mixed modes, we found
posterior probabilities of 0.001\% and 8.7\%, respectively. These peaks are therefore probably due to signal. 
The larger posterior probability found for $\pi_4$ is caused by the fact that the expected profile of mixed modes was built with very conservative priors . 
For $\pi_2$ and $\pi_3$, assumed to correspond to $\ell=3$ modes, we 
obtained a very low posterior probability (1.4\% and 1.1\%, respectively), indicating that these peaks are highly likely due to signal, and a good compatibility
between their observed profiles and the expected profile of $\ell=3$ modes for \cible. This confirms our identification for $\pi_2$ and $\pi_3$.
It is the first time that $\ell=3$ modes can be detected in a solar-like pulsator (other than the Sun). We explain in Sect. \ref{sect_mode_param} why \cible\
was a particularly favorable target to detect $\ell=3$ modes.

We remark that ridge D could also be seen as the $m=+1$ component of the rotationally
splitted $\ell=1$ mode. This hypothesis is studied in \ref{sect_alternate}. 

\subsection{Extraction of p mode parameters \label{sect_fit}}

\subsubsection{Method \label{sect_method}}

Solar-like oscillations are the result of stochastic excitations of p modes in the convective envelope. They respond to the equation of a forced, damped and randomly excitated harmonic oscillator (see 
\citealt{1986ssds.proc..105D}). Therefore their signature in the power spectrum can be modeled as Lorentzian profiles with a linewidth proportional to the inverse of the mode lifetimes,
added to the background and perturbated by a multiplicative noise following a $\chi^2$ distribution with two degrees of freedom. We used a maximum 
likelihood estimation (MLE) to determine the most probable parameters of the observed p modes (see \citealt{1990ApJ...364..699A}). This method was already
applied to determine p-mode parameters for the Sun (\citealt{Toutain92}, \citealt{Appourchaux98}) as well as for \corot\ solar-like targets 
(\citealt{2008A&A...488..705A}, \citealt{HD181420}, \citealt{2009A&A...506...41G}).
As described in \cite{1990ApJ...364..699A}, the likelihood $L$ is defined as the product of the probability density of each frequency bin $\nu_i$ of the power spectrum
\begin{equation}
L(\nu_i,\boldsymbol{\lambda})=\prod_{i}\frac{1}{M(\nu_i,\boldsymbol{\lambda})}\exp \left( \frac{-S(\nu_i)}{M(\nu_i,\boldsymbol{\lambda})} \right),
\end{equation}
where $S$ is the observed power spectrum, $M$ the model described above, and $\boldsymbol{\lambda}$ the parameters of the Lorentzian profiles
used in the model. The set of parameters $\boldsymbol{\lambda}$ that maximizes the likelihood corresponds to the most probable parameters. Maximizing $L$ is 
the same as minimizing the negative logarithmic likelihood $\mathcal{L}$, defined as
\begin{equation}
\mathcal{L}\equiv-\ln(L)=\sum_i \left( \ln M(\nu_i,\boldsymbol{\lambda}) + \frac{S(\nu_i)}{M(\nu_i,\boldsymbol{\lambda})} \right).
\end{equation}

As mentioned above, perturbations remain in the power spectrum at frequencies corresponding to combinations of the harmonics of the orbital frequency and those of the day.
Some of these perturbations can be seen in the frequency domain of the oscillations. We therefore gave a null weight to the corresponding bins so that our estimates of the p-mode parameters
are not influenced by those peaks.

We estimated the background with a power law to describe the granulation component, as prescribed by \cite{1985ESASP.235..199H}. Prior to the fit of the p-mode parameters,
we fitted a profile of the type
\begin{equation}
B(\nu_i)=\frac{\alpha}{1+(\beta\nu_i)^{\gamma}}+\delta
\end{equation}
on the power spectrum, excluding the frequency domain of the p-mode oscillations to avoid any bias, with $\alpha,\beta,\gamma$ and $\delta$ as free parameters. 
We then used this background to fit the mode parameters.

The analysis of HD~49933, which was the first solar-like \corot\ target, showed that a global fit of all the modes gave more robust estimates of the mode parameters
(see \citealt{2008A&A...488..705A}). We therefore fitted modes simultaneously on a frequency range corresponding to nine overtones in the power spectrum. 
The assumptions made for the fit are the same as in \cite{2008A&A...488..705A}. We briefly recall them here.
The splitting was assumed to be constant over the frequency range of the oscillations, and the $m\neq0$ components of non-radial modes were supposed to be symmetrical
with respect to the $m=0$ component.
We sliced the power spectrum in $\Delta\nu$-wide intervals, in which the linewidth is assumed to be the same for all modes, and where only the height of the $\ell=0$ mode ($H_{\ell=0,n}$)
is fitted. The heights of non-radial modes are determined from those of the $\ell=0$ mode. Assuming an equal repartition of the energy in the modes of different degrees $\ell$, and taking
the value of the limb darkening for this star into account, we obtain
\begin{eqnarray*}
H_{\ell=1,n}/H_{\ell=0,n} & = & 1.5 \\
H_{\ell=2,n}/H_{\ell=0,n} & = & 0.5 \\
H_{\ell=3,n}/H_{\ell=0,n} & = & 0.05,
\end{eqnarray*}
where $H_{\ell=1,n}=\sum_{m=-\ell}^{m=+\ell} H_{n,\ell,m}$.
Inside non-radial multiplets, the heights $H_{n,\ell,m}$ are given by the intensity visibilities
(see \citealt{2003ApJ...589.1009G}, \citealt{2006MNRAS.369.1281B}).

\subsubsection{Mode parameters estimates \label{sect_mode_param}}

Five teams worked independently on the extraction of the p-mode parameters and applied the method we described in Sect. \ref{sect_method} to fit the four ridges 
associated to $\ell=0,1,2,3$ modes. As was done in previous analyses of this type, we present in the following section
one set of reference results, and we use results from the other teams to cross-check them and comment the robustness of the 
solution. The reference results of the fit are given in Tables \ref{tab_param0A} and \ref{tab_param12A}.

\begin{table*}
\centering
\caption{Reference results for the parameters of the radial p modes. The heights given here are those of the $\ell=0$ modes (the heights of non-radial modes are computed using height ratios, 
see Sect. \ref{sect_method}). The agreement (agr.) between the teams is given for each fitted parameter. Filled circles indicate that at least four of the five teams agreed on the value within 1 $\sigma$. 
$X\,\sigma$ indicates that more than one team disagrees with the value, and that the five teams agree within $X\,\sigma$ ($X>1$).
\label{tab_param0A}}
%\begin{tabular}{| c | c c | c c | c c | c c | c |}
\begin{tabular}{ l c c c c c c c c c }
\hline \hline
$\ell$ \T \B & $\nu_{n,\ell}$ ($\mu$Hz) & agr.($\nu_{n,\ell}$) & $H_{0,n}$ (ppm$^2/\mu$Hz) & agr.($H_{0,n}$) & $\Gamma_n$ ($\mu$Hz) & agr.($\Gamma_n$) & $a_n$ (ppm) & agr.($a_n$) & $H_{\ell,n}/B(\nu_{n,\ell})$ \\
\hline
0 \T \B & $ 799.70\pm 0.27$ & $\bullet$ & $2.55^{+0.80}_{-0.61}$ & $\bullet$ & $1.58^{+0.42}_{-0.33}$ & $\bullet$ & $3.55^{+0.67}_{-0.56}$ & $\bullet$  & 2.23 \\
0 \T \B& $ 855.30\pm 0.73$ & $\bullet$ & $2.10^{+0.59}_{-0.46}$ & $\bullet$ & $2.97^{+0.72}_{-0.58}$ & $\bullet$ & $4.42^{+0.77}_{-0.65}$ & $\bullet$  & 1.94 \\
0 \T \B& $ 909.92\pm 0.26$ & $\bullet$ & $3.13^{+0.82}_{-0.65}$ & $\bullet$ & $2.00^{+0.43}_{-0.35}$ & $\bullet$ & $4.44^{+0.70}_{-0.60}$ & $\bullet$  & 3.03 \\
0 \T & $ 965.16\pm 0.36$ & $\bullet$ & $4.05^{+0.96}_{-0.78}$ & $\bullet$ & $2.37^{+0.43}_{-0.37}$ & $\bullet$ & $5.49^{+0.83}_{-0.72}$ & $\bullet$  & 4.09 \\
0 \T \B& $1021.81\pm 0.28$ & $\bullet$ & $4.56^{+1.15}_{-0.92}$ & $\bullet$ & $2.03^{+0.41}_{-0.34}$ & $\bullet$ & $5.40^{+0.80}_{-0.70}$ & $\bullet$  & 4.78 \\
0 \T \B& $1078.97\pm 0.33$ & $\bullet$ & $4.77^{+1.10}_{-0.90}$ & $\bullet$ & $2.08^{+0.35}_{-0.30}$ & $\bullet$ & $5.58^{+0.80}_{-0.70}$ & $\bullet$  & 5.18 \\
0 \T \B& $1135.32\pm 0.34$ & $\bullet$ & $3.49^{+0.83}_{-0.67}$ & $\bullet$ & $2.37^{+0.44}_{-0.37}$ & $\bullet$ & $5.09^{+0.86}_{-0.73}$ & $\bullet$  & 3.90 \\
0 \T \B& $1192.12\pm 0.45$ & $\bullet$ & $2.29^{+0.67}_{-0.52}$ & $\bullet$ & $2.51^{+0.68}_{-0.53}$ & $1.8\,\sigma$ & $4.25^{+0.80}_{-0.67}$ & $\bullet$  & 2.63 \\
0 \T \B& $1250.12\pm 0.89$ & $\bullet$ & $1.15^{+0.32}_{-0.25}$ & $\bullet$ & $3.90^{+1.00}_{-0.80}$ & $\bullet$ & $3.76^{+4.04}_{-1.95}$ & $\bullet$  & 1.36 \\
\hline
\end{tabular}
\end{table*}

\begin{figure}
\centering
\includegraphics[width=8.5cm]{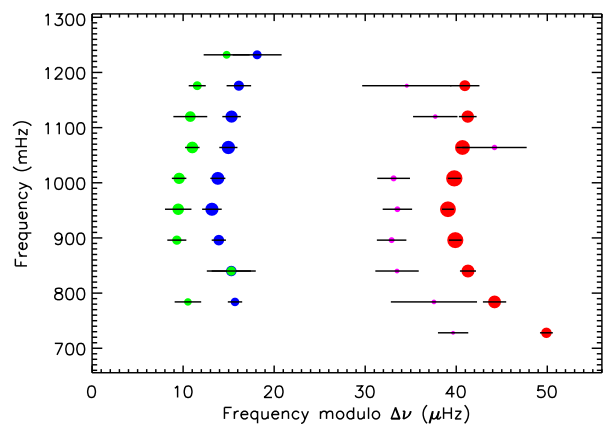}
\caption{\'Echelle diagram of the fitted modes. The colors correspond to the following degrees: blue $\ell=0$, red $\ell=1$, green $\ell=2$ and purple $\ell=3$. The surface of the disks
is proportional to the heights of the modes. 3-$\sigma$ error bars are overplotted.
\label{fig_ech_fitA}}
\end{figure}

\paragraph{Mode frequencies}
~~\\\\
There is a very good agreement between the different teams on the fitted mode eigenfrequencies. Only two frequencies in the $\ell=3$ ridge 
differ by more than 1 $\sigma$. Indeed, this ridge is too faint to offer reliable estimates of the mode frequencies away from the maximum of the signal. 
An \'echelle diagram of the fitted modes is given in Fig. \ref{fig_ech_fitA}.
At the maximum of the signal, the obtained precision on the mode frequencies ranges from about 0.2 to about 0.6 $\mu$Hz for $0\leqslant\ell\leqslant3$ modes
and increases at the edges of the frequency domain of the oscillations.

At $\nu=855.3\,\mu$Hz, the $\ell=2$ mode is found merged together with the $\ell=0$ mode. This could be because the signal-to-noise ratio decreases
when reaching the edges of the fitted domain. Indeed, with a lower signal-to-noise ratio, the analysis of previous \corot\ targets yielded several $\ell=2$ modes merged with
the neighboring $\ell=0$ mode (see \citealt{2008A&A...488..705A}, \citealt{HD181420}). However, the $\ell=2$ mode might also have been shifted because of
an $\ell=2$ mixed mode in the neighborhood.

\begin{figure}
\centering
\includegraphics[width=8.5cm]{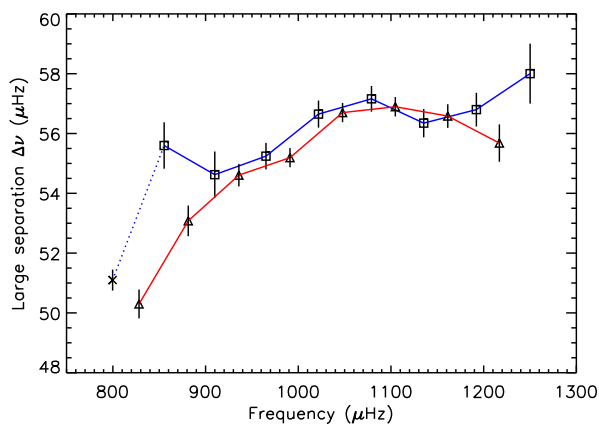}
\caption{Large separation profiles derived from the fitted frequencies of the $\ell=0$ (squares and blue line) and 
$\ell=1$ (triangles and red line) modes. We overplotted the 1-$\sigma$ error bars. The cross corresponds to the 
large separation we would obtain, if $\pi_1$ were considered as an $\ell=0$ mode.
\label{fig_large_sep}}
\end{figure}

Figure \ref{fig_large_sep} shows the large separation profiles derived from the fitted frequencies of the $\ell=0$ and $\ell=1$ modes. 
It appears, as mentioned in Sect. \ref{sect_id}, that the $\ell=0$ and $\ell=1$ ridges have a different curvature at low frequency, 
which might be caused by a low-frequency avoided crossing.

\paragraph{Mode linewidths}
~~\\\\

\begin{figure}
\centering
\includegraphics[width=8.5cm]{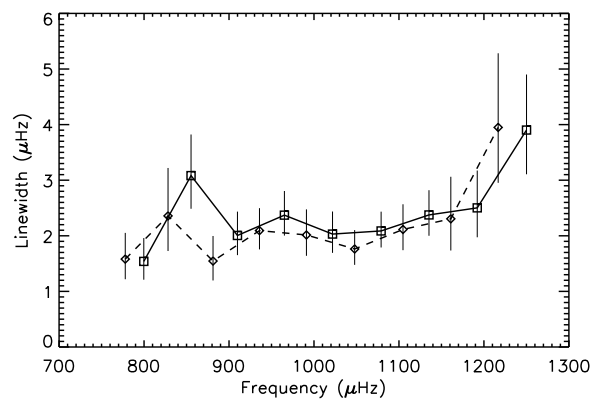}
\caption{Linewidths of the fitted modes with 1-$\sigma$ error bars (squares and full line). The linewidths for the alternate identification are also given (diamonds and dashed line).
\label{fig_larg}}
\end{figure}

Figure \ref{fig_larg} shows the linewidths of the fitted modes. As was observed in the Sun, the linewidths of the modes increase with frequency and are almost constant around the maximum
of the signal. They are larger than those of the Sun by a factor 2. 
%We thus obtain an estimate of the mode lifetimes for a cooler and more evolved object than the solar-like pulsators previously 
%observed with \corot\ and for which mode linewidhts could be estimated. 
The lifetimes $\tau$ of the modes can be deduced from the linewidth by the relation $\tau=1/(\pi\Gamma)$.
For \cible\ they range from about one day to two days. By comparison the mode lifetimes are shorter in other \corot\ solar-like pulsators,
\textit{e.g.} for HD\,181420 (about six hours for the shortest ones, and 12 hours around the maximum of the signal).

It appears that one of the fitted intervals ($\nu=855.3\,\mu$Hz for the $\ell=0$ mode) has a linewidth somewhat larger than what could be expected considering 
the linewidths of neighboring orders (see Fig. \ref{fig_larg}). Indeed, the $\ell=1$ mode of this interval ($\nu=828.2\,\mu$Hz) is larger than the other modes in the neighborhood.
%It could possibly be due to the presence of a mixed mode with an eigenfrequency close to that of the $\ell=1$ mode, and which broadens the fitted peak.

\paragraph{Mode heights and amplitudes}
~~\\\\

As was mentioned before, we fixed the ratios of heights beween modes of different degrees $\ell$ in our fit. Because \cible\ is the first \corot\ target for which the $\ell=0$ and $\ell=2$ ridges
are clearly distinct in the \'echelle diagram, we had the opportunity to check the chosen height ratios, which are those usually adopted in this type of analysis. 
We performed a new fit releasing the height ratios and found
\begin{eqnarray*}
H_{\ell=1,n}/H_{\ell=0,n} & = 1.54\pm0.16  \\
H_{\ell=2,n}/H_{\ell=0,n} & = 0.76\pm0.12  \\
H_{\ell=3,n}/H_{\ell=0,n} & = 0.20\pm0.05.
\end{eqnarray*}
For $\ell=1$ modes, the theoretical value of the ratio lies within the 1-$\sigma$ error bars of the fitted value. For $\ell=2$ and $\ell=3$ modes, the theoretical ratios 
are at 2 and 3 $\sigma$ of the fitted values, respectively. At this stage, this difference is too small to be interpreted as an inconsistency with the theoretical
height ratios. It would be interesting however to investigate this matter in future analyses of this type on different objects. 
We noticed that with free height ratios the results for mode frequencies, linewidths and amplitudes remained within
the 1-$\sigma$ error bars of the parameters obtained with fixed height ratios.

\begin{figure}
\centering
\includegraphics[width=8.5cm]{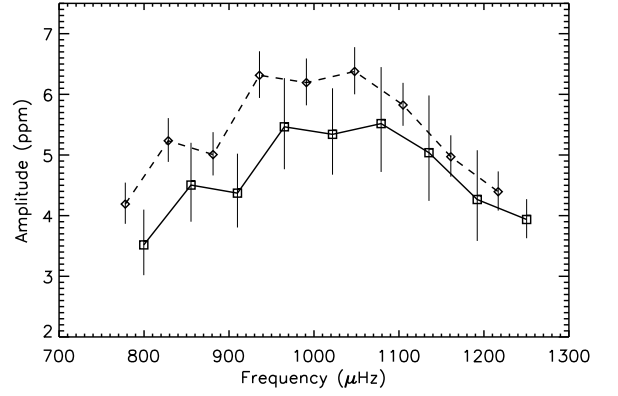}
\caption{Amplitudes of the fitted modes with 1-$\sigma$ error bars (squares and full line). The amplitudes for the alternate identification also appear (diamonds and dashed line).
\label{fig_amp}}
\end{figure}

From the heights and linewidths of the modes we derived the amplitudes (in ppm), given by $A=\sqrt{\pi H \Gamma}$. To estimate the error bars on this quantity,
we took into account the anticorrelation which exists between the determination of the height and the linewidth. The mode amplitudes are represented for $\ell=0$ modes as a function
of frequency in Fig. \ref{fig_amp}. They range from 3.6 ppm to 5.4 ppm. However, these estimates correspond to instrumental values. Taking into account the instrumental response 
functions of \corot\ (as described in \citealt{michel09}), they can be converted into intrinsic bolometric amplitudes per radial mode by multiplying them by a factor 1.032.
The obtained amplitudes are consistent with the prediction of the maximum amplitudes using scaling laws ($A_{\ell=0}=6.1\pm0.5$, see Appendix \ref{app_amp}).
The star \cible\ is then quite specific among the other solar-like pulsators observed with \corot\ for which amplitudes were found significantly lower than the predicted ones
(see \citealt{2008A&A...488..705A}, \citealt{HD181420}, \citealt{2009A&A...506...41G}, \citealt{2008Sci...322..558M}).

The signal-to-noise (SNR) ratio of each mode computed from the fitted height and the background estimated in Sect. \ref{sect_method} is given in Table \ref{tab_param0A}.
The SNR is found to be very small for the fitted $\ell=3$ modes (about 0.25 at the maximum of the signal). However, when releasing the height ratios, we find
$H_{\ell=3,n}/H_{\ell=0,n} = 0.20\pm0.05$. The SNR of the $\ell=3$ modes should be closer to 1 at the maximum of the signal.

\begin{table}
\centering
\caption{Reference results for the parameters of the non-radial p modes (symbols are the same as in Table \ref{tab_param0A}).
\label{tab_param12A}}
%\begin{tabular}{| c | c c | c c | c c | c c | c |}
\begin{tabular}{ l c c c}
\hline \hline
$\ell$ \T \B & $\nu_{n,\ell}$ ($\mu$Hz) & agr.($\nu_{n,\ell}$) & $H_{\ell,n}/B(\nu_{n,\ell})$ \\
\hline
1 \T \B& $ 777.91\pm 0.24$ & $\bullet$  & 3.26 \\
1  \B& $ 828.21\pm 0.42$ & $\bullet$  &  2.83 \\
1  \B& $ 881.29\pm 0.29$ & $\bullet$  & 4.44 \\
1  \B& $ 935.90\pm 0.23$ & $\bullet$  &  6.00 \\
1  \B& $ 991.09\pm 0.22$ & $\bullet$  &  7.03 \\
1  \B& $1047.79\pm 0.24$ & $\bullet$  & 7.62 \\
1  \B& $1104.68\pm 0.22$ & $\bullet$  &  5.76 \\
1  \B& $1161.27\pm 0.33$ & $\bullet$  &  3.89 \\
1  \B& $1216.95\pm 0.53$ & $\bullet$  &  2.01 \\
\hline
2  \T \B& $ 794.55\pm 0.52$ & $\bullet$  &  1.11 \\
2  \B& $ 855.29\pm 0.93$ & $\bullet$  &  0.97 \\
2  \B& $ 905.31\pm 0.35$ & $\bullet$  &  1.51 \\
2  \B& $ 961.47\pm 0.49$ & $\bullet$  &  2.04 \\
2  \B& $1017.56\pm 0.27$ & $\bullet$  & 2.38 \\
2  \B& $1075.01\pm 0.27$ & $\bullet$  &  2.58 \\
2  \B& $1130.79\pm 0.61$ & $\bullet$  & 1.95 \\
2  \B& $1187.55\pm 0.32$ & $\bullet$  &  1.31 \\
2  \B& $1246.78\pm 0.84$ & $\bullet$  &  0.68 \\
\hline
3 \T \B & $ 767.74\pm 0.65$ & $7.3\,\sigma$  & 0.11 \\
3 \B & $ 821.55\pm 1.57$ & $\bullet$   & 0.09 \\
3 \B & $ 873.46\pm 0.83$ & $\bullet$  &  0.15 \\
3 \B & $ 928.93\pm 0.57$ & $\bullet$  &  0.20 \\
3 \B & $ 985.53\pm 0.55$ & $\bullet$  & 0.23 \\
3 \B & $1041.11\pm 0.59$ & $\bullet$  &  0.25 \\
3 \B & $1108.11\pm 1.33$ & $3.1\,\sigma$  &  0.19 \\
3 \B & $1157.70\pm 0.83$ & $\bullet$  &  0.13 \\
3 \B & $1210.55\pm 1.68$ & $\bullet$  &  0.07 \\
\hline
\end{tabular}
\end{table}

\paragraph{Splitting and inclination angle estimates }
~~\\\\
Most of the fits we performed converge toward a null value of the splitting
and consequently an undefined value of the inclination angle. This can correspond to two different cases:
\begin{itemize}
\item The star could be seen almost pole-on, i.e. with an inclination angle close to $0^{\circ}$. In that case the $m\neq0$ components of the non-radial 
modes are not visible, and we therefore cannot estimate the splitting. However, the measured $v\sin i$ of the object and the estimate of the stellar radius $R$ allow us to give constraints
on the splitting $\nu\ind{s}$ through the relation
\begin{equation}
v \sin i = 2\pi R \nu\ind{s} \sin i.
\label{eq_vsini}
\end{equation}
For a low inclination angle we expect higher values of the rotational splitting (\textit{e.g.} $\nu\ind{s}>1.5\,\mu$Hz for $i$ lower than $10^{\circ}$).
\item We could also explain this result by a low value of the splitting compared to the linewidth of the modes, so that the fits are unable to separate the components
of the non-radial modes. Equation \ref{eq_vsini} shows that a small rotational splitting corresponds to a high inclination angle.
%Some of our fits converge toward an angle close to $90^\circ$ and a small value of the splitting.
%Using Eq. \ref{eq_vsini}, this solution corresponds to a splitting of about $\nu\ind{s}\simeq 0.3\,\mu$Hz.
\end{itemize}
This fitting procedure does not significantly favor any of these two hypotheses. However, they lead to almost identical p mode parameters (all the parameters agree within 1-$\sigma$ error bars).

In both cases, the effect of the rotational splitting on the observed mode profiles is weak. 
Indeed, either the angle is small, and the $m\ne0$ components of the multiplets have very small amplitudes compared
to the $m=0$ component, or the splitting is small, and the $m$ components are mixed together. We note that this favors the detection of lower degree modes and therefore justifies
why $\ell=3$ modes seem to be detected in \cible, while they were not in previously analyzed \corot\ targets.

\subsubsection{Possible mixed modes parameters estimates \label{sect_mixed}}

The position of \cible\ in the HR diagram suggests that it either reaches the end of the main sequence or lies shortly after it. 
The star \cible\ is therefore evolved and its
acoustic spectrum may contain mixed modes, i.e. modes which have a g-mode behavior in the deep 
interior and a p-mode behavior below the surface (see \citealt{1974A&A....36..107S}). 
The mixed modes are known to experience the so-called avoided crossing phenomenon, during which their frequency 
does not follow the asymptotic approximation and they can be located out of the ridges (\citealt{unno89}). 
It was already noticed (\textit{e.g.} \citealt{1991A&A...248L..11D}) that the detection of these modes 
could yield a very valuable constraint on the size of the convective core. 

As was noticed in Sect. \ref{sect_ls}, two peaks are detected in the power spectrum with a confidence level above 95\% and are not located
within any of the observed ridges in the \'echelle diagram (namely $\pi_1$ and $\pi_4$).
They could be the signature of mixed modes. We note that if the peak $\pi_1$, found at $\nu\sim748\,\mu$Hz is an $\ell=1$ mixed
mode in avoided crossing, it could explain why the $\ell=1$ p mode at $\nu\sim797.5\,\mu$Hz seems not to follow the curvature of the $\ell=1$ ridge.
We fitted these peaks as Lorentzian profiles with free frequencies, heights and linewidths. Because the mixed modes have a g-mode character near the center,
their inertia is larger, and we expect them to have a longer lifetime and therefore a smaller linewidth than regular modes.
We obtained the results given in Table \ref{tab_mixed}. We notice that the fitted frequencies have small error bars, but the other parameters are ill-determined.

\begin{table}
  \centering
  \caption{Results of a Lorentzian fit performed on the peaks identified as potential mixed modes.
  \label{tab_mixed}}
\begin{tabular}{c  c  c  c }
\hline \hline
$\nu$ ($\mu$Hz) & $H$ (ppm$^2/\mu$Hz) & $\Gamma$ ($\mu$Hz) & $a$ (ppm) \\
\hline
\T \B $748.60\pm0.23$ & $2.82^{+2.08}_{-1.20}$ & $1.12^{+0.79}_{-0.46}$ & $3.15_{-0.92}^{+1.32}$ \\
%$1002.39\pm0.23$ & 3.64 $+11.73$/$-2.78$ & 0.29 $+1.43$/$-0.24$ & 1.82 $+2.92$/$-1.12$ \\
\T \B $1201.96\pm0.23$ & $2.03^{+2.05}_{-1.02}$ & $0.67^{+0.57}_{-0.31}$ & $2.07^{+1.06}_{-0.70}$ \\
\hline
\end{tabular}
\end{table}

\subsubsection{Other mode identification scenarii \label{sect_alternate}}

\begin{table*}
\centering
\caption{Reference results for the parameters of the radial p modes for the \textbf{alternate identification} (symbols are the same as in Table \ref{tab_param0A}).
\label{tab_param0B}}
\begin{tabular}{ l c c c c c c c c c }
\hline \hline
$\ell$ \T \B & $\nu_{n,\ell}$ ($\mu$Hz) & agr.($\nu_{n,\ell}$) & $H_{0,n}$ (ppm$^2/\mu$Hz) & agr.($H_{0,n}$) & $\Gamma_n$ ($\mu$Hz) & agr.($\Gamma_n$) & $a_n$ (ppm) & agr.($a_n$) & $H_{\ell,n}/B(\nu_{n,\ell})$ \\
\hline
0 \T \B & $ 777.89\pm 0.24$ & $\bullet$ & $3.23^{+1.02}_{-0.78}$ & $\bullet$ & $1.58^{+0.47}_{-0.36}$ & $\bullet$ & $4.00^{+0.31}_{-0.29}$ & $\bullet$  & 2.76 \\
0 \T \B & $ 828.22\pm 0.42$ & $\bullet$ & $3.17^{+1.16}_{-0.85}$ & $\bullet$ & $2.36^{+0.86}_{-0.63}$ & $\bullet$ & $4.85^{+0.29}_{-0.28}$ & $1.4\,\sigma$  & 2.85 \\
0 \T \B & $ 881.18\pm 0.34$ & $\bullet$ & $4.53^{+1.42}_{-1.08}$ & $\bullet$ & $1.55^{+0.45}_{-0.35}$ & $\bullet$ & $4.69^{+0.29}_{-0.27}$ & $\bullet$  & 4.28 \\
0 \T \B & $ 935.83\pm 0.22$ & $\bullet$ & $5.32^{+1.17}_{-0.96}$ & $\bullet$ & $2.09^{+0.40}_{-0.34}$ & $\bullet$ & $5.92^{+0.28}_{-0.27}$ & $\bullet$  & 5.26 \\
0 \T \B & $ 991.09\pm 0.25$ & $\bullet$ & $5.32^{+1.30}_{-1.05}$ & $\bullet$ & $2.01^{+0.46}_{-0.37}$ & $\bullet$ & $5.80^{+0.29}_{-0.28}$ & $\bullet$  & 5.47 \\
0 \T \B & $1047.86\pm 0.29$ & $\bullet$ & $6.45^{+1.52}_{-1.23}$ & $\bullet$ & $1.76^{+0.34}_{-0.29}$ & $\bullet$ & $5.98^{+0.29}_{-0.27}$ & $\bullet$  & 6.88 \\
0 \T \B & $1104.74\pm 0.22$ & $\bullet$ & $4.49^{+1.06}_{-0.86}$ & $\bullet$ & $2.11^{+0.45}_{-0.37}$ & $\bullet$ & $5.46^{+0.27}_{-0.26}$ & $\bullet$  & 4.95 \\
0 \T \B & $1161.01\pm 0.34$ & $\bullet$ & $2.99^{+0.96}_{-0.73}$ & $\bullet$ & $2.31^{+0.76}_{-0.57}$ & $\bullet$ & $4.65^{+0.27}_{-0.26}$ & $\bullet$  & 3.38 \\
0 \T  \B & $1216.77\pm 0.69$ & $\bullet$ & $1.38^{+0.42}_{-0.32}$ & $\bullet$ & $3.95^{+1.33}_{-1.00}$ & $\bullet$ & $4.14^{+0.28}_{-0.26}$ & $\bullet$  & 1.60 \\
\hline
\end{tabular}
\end{table*}

\paragraph{Can ridge D be the $m=+1$ component of the $\ell=1$ modes?}
~~\\\\
Still considering the identification preferred in this paper (\textit{i.e.} ridge C corresponding to $\ell=1$ modes),
ridge D could also possibly be seen as the $m=+1$ of the $\ell=1$ ridge, as already mentioned in Sect. \ref{sect_ls}. When not including the $\ell=3$ modes in the fit, we indeed
find solutions with a splitting of about 7 $\mu$Hz, corresponding to the distance between ridges C and D. 

However, we ran models representative of \cible\
and found that with this rotation frequency the $m\neq0$ components of 
non-radial multiplets are no longer symmetric with respect to the $m=0$ component.
To treat this case correctly, we therefore need to take this asymmetry into account in our fits. 

\begin{figure}
\centering
\includegraphics[width=8.5cm]{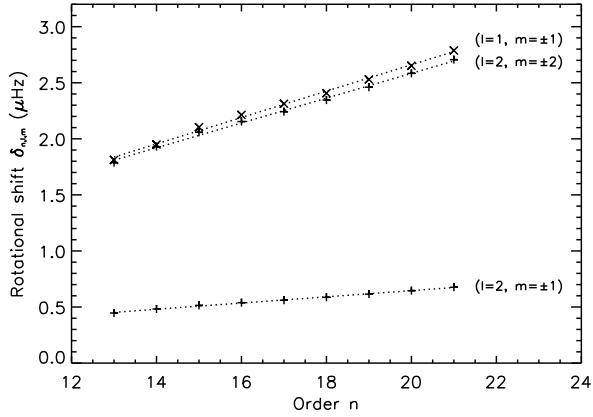}
\caption{Frequency shift $\rotshift$ (defined in Eq. \ref{eq_shift}) of the $m\neq0$ components of non-radial multiplets for an evolutionary model representative of \cible\ in the
frequency range of the detected oscillations and with a splitting of 3 $\mu$Hz. Crosses stand for $\ell=1$ modes and plus signs for $\ell=2$ modes. 
A linear regression of the variations of the shifts with respect to the frequency is overplotted (dotted lines).
\label{fig_asym}}
\end{figure}

For this purpose, we studied this asymmetry on evolutionary models reproducing
the $T\ind{eff}$ and $L$ that we obtained for \cible\ in Sect. \ref{sect_teff} and with different angular velocities. 
The models were computed with the evolutionary code CESAM2k (\citealt{1997A&AS..124..597M}) 
and the eigenfrequencies were derived from them with the oscillation code FILOU (\citealt{2008Ap&SS.316..155S}). 
They showed that the effects of asymmetry should be non-negligible even for values of the splitting as low as a few $\mu$Hz.
We describe the asymmetry of the $m\neq0$ components of a mode of degree $\ell$ and order $n$ by the shift $\rotshift$ defined as
\begin{equation}
\delta_{n,\ell,m}\equiv\nu_{n,\ell,0}-\frac{\nu_{n,\ell,+m}+\nu_{n,\ell,-m}}{2}.
\label{eq_shift}
\end{equation}
These shifts are represented in Fig. \ref{fig_asym} for a model representative of \cible\ for eigenmodes in the frequency range of the detected oscillations ($13\leqslant n\leqslant21$)
and with a rotational splitting of 3 $\mu$Hz. 
We observe that the shifts $\rotshift$ are expected to vary approximately linearly with respect to frequency. We therefore model the shift $\rotshift$ as
\begin{equation}
\rotshift=\alpha_{\ell,m}(n-n_0)+\delta_{n_0,\ell,m},
\end{equation}
where $n_0$ corresponds to the order where the signal is maximal ($n_0=17$). For $m\neq0$ modes in non-radial multiplets we now fit eigenfrequencies of the type
\begin{equation}
\nu_{n,\ell,m}=\nu_{n,\ell,0}+m\nu\ind{s}-\rotshift.
\end{equation}
This adds two free parameters to the fit per fitted $m\neq0$ component.

The obtained results for the eigenfrequencies, heights, linewidths and amplitudes of modes of degrees $0\leqslant\ell\leqslant2$ all lie within the 1-$\sigma$ error bars of those
given in Table \ref{tab_param0A}. The fit converges toward a splitting of $\nu\ind{s}=3.8\pm0.2\,\mu$Hz and an angle of $i=20.5\pm2.3^{\circ}$.
As expected, ridge D is fitted as the $m=+1$ component of the $\ell=1$ modes.
The obtained rotational shift is such that $\alpha_{\ell=1,m=1}=0.30\pm0.19\,\mu$Hz per order $n$, and $\delta_{n_0,\ell=1,m=1}=3.0\pm0.6\,\mu$Hz.
However, using Eq. \ref{eq_vsini} and the fitted values of the splitting and the angle, we obtain $v\sin i=11.2\pm2.8$ \kms.
This result is not compatible with the firm upper limit of 5 \kms\ established for \vsini\ in the spectroscopic analysis (see Sect. \ref{sect_vsini}).

\paragraph{Can ridge C correspond to $\ell=0$ modes?}
~~\\\\
As was said in Sect. \ref{sect_ls}, for the sake of completeness and in the light of previous analyses of p-mode oscillations in solar-like pulsators, we decided to also investigate the
alternate identification. In this case, ridges C and D correspond to $\ell=0$ and $\ell=2$ modes, respectively. Ridge A could not possibly be an $\ell=3$ ridge, because it has much
too strong amplitudes. Ridges A and B would then be the $m=\pm1$ components of rotationally splitted $\ell=1$ modes.

\begin{figure}
\centering
\includegraphics[width=8.5cm]{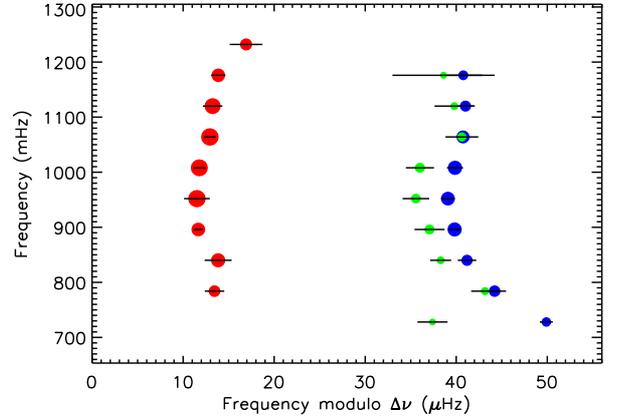}
\caption{\'Echelle diagram for the fitted modes of the \textbf{alternate identification}. Same symbols as in Fig. \ref{fig_ech_fitA}.
\label{fig_ech_fitB}}
\end{figure}

We fitted the modes exactly the same way it was done before. The results are given in Tables \ref{tab_param0B} and \ref{tab_param12B}, and an \'echelle diagram of the fitted
modes is shown in Fig. \ref{fig_ech_fitB}. The linewidths, heights and amplitudes are very comparable with those obtained in the first identification. We obtain a splitting of
$\nu\ind{s}=2.2\pm0.1\,\mu$Hz and an angle of $i=70.6\pm5.8^{\circ}$, which is consistent with the idea that ridges A and B are indeed identified as $(\ell=1,m=\pm1)$ modes because they
are separated by about twice the fitted splitting. Using Eq. \ref{eq_vsini} as before, these fitted values of the splitting and angle give an estimate of $v\sin i=17.7\pm2.8$ \kms, in complete
disagreement with the measured $(v\sin i)\ind{measured}=2.9^{+1.0}_{-1.5}$ \kms. \\

The alternate identifications we considered lead to major disagreements with the measured \vsini\ and were therefore rejected. This confirms the identification 
we adopted in Sect. \ref{sect_id} to perform the global fit of the p-mode parameters.

\begin{table}
\centering
\caption{Reference results for the parameters of the non-radial p modes for the \textbf{alternate identification} (symbols are the same as in Table \ref{tab_param0A}).
\label{tab_param12B}}
\begin{tabular}{ l c c c }
\hline \hline
$\ell$ \T \B & $\nu_{n,\ell}$ ($\mu$Hz) & agr.($\nu_{n,\ell}$) & $H_{\ell,n}/B(\nu_{n,\ell})$ \\
\hline
1 \T \B& $ 797.45\pm 0.36$ & $46.0\,\sigma$  &  4.22 \\
1  \B& $ 853.85\pm 0.49$ & $\bullet$  &  4.39 \\
1  \B& $ 907.67\pm 0.18$ & $\bullet$  &  6.56 \\
1  \B& $ 963.52\pm 0.47$ & $\bullet$  &  8.05 \\
1  \B& $1019.78\pm 0.22$ & $\bullet$  &  8.36 \\
1  \B& $1076.94\pm 0.22$ & $\bullet$  &  10.49 \\
1  \B& $1133.24\pm 0.36$ & $\bullet$  &  7.53 \\
1  \B& $1189.87\pm 0.26$ & $\bullet$  &  5.15 \\
1  \B& $1248.92\pm 0.60$ & $\bullet$  &  2.44 \\
\hline
2 \T \B & $ 765.39\pm 0.55$ & $39.7\,\sigma$  &  1.36 \\
2 \B & $ 827.17\pm 0.51$ & $27.2\,\sigma$  &  1.43 \\
2 \B & $ 878.29\pm 0.38$ & $\bullet$  & 2.13 \\
2 \B & $ 933.07\pm 0.55$ & $\bullet$  &  2.62 \\
2 \B & $ 987.57\pm 0.49$ & $\bullet$  &  2.73 \\
2 \B & $1044.01\pm 0.52$ & $\bullet$ &  3.43 \\
2 \B & $1104.62\pm 0.60$ & $\bullet$  & 2.47 \\
2 \B & $1159.79\pm 0.72$ & $\bullet$  & 1.69 \\
2 \B & $1214.61\pm 1.87$ & $\bullet$  &  0.80 \\
\hline
\end{tabular}
\end{table}

\section{Conclusion\label{sect_concl}}

The star HD~49385 was characterized from both \corot\ seismic observations and NARVAL spectroscopic data.
The atmospheric parameters of the star were derived and the 137-day-long photometric time series was analyzed.

A clear series of peaks associated to p modes was detected in the power spectrum around 1 mHz.
Up to now, the \corot\ solar-like pulsators presented some ambiguities in the identification of ridges corresponding to different degrees $\ell$.
Here three very clear ridges appear in the \'echelle diagram, which were readily identified as $\ell=0$, 1 and 2
modes. The $\ell=2$ ridge appears clearly distinct from the $\ell=0$ ridge in the \'echelle diagram.
Furthermore, two peaks, part of a fainter ridge, were found to be significant and compatible with the
characteristics we expect for $\ell=3$ modes in \cible. This probably constitutes the first photometric detection of $\ell=3$ modes in a
solar-like pulsating star (other than the Sun).

We performed a global fit over nine radial orders with degrees ranging from $\ell=0$ to $\ell=3$ modes (simultaneous
fit of 36 individual modes). We obtained precise estimates of the mode frequencies (uncertainty of about $0.2\,\mu$Hz for radial modes at the maximum of the signal).
The obtained value of the maximum amplitude for $\ell=0$ modes ($5.6\pm0.8$ ppm) is consistent
with the estimate deduced from \cite{samadi07}, unlike the other \corot\ solar-like pulsators (\textit{e.g.} \citealt{2008A&A...488..705A}).

We found no evidence of a rotational splitting of the modes. This can be explained either by a small inclination angle or by a low rotational velocity
(inducing a small rotational splitting compared to the linewidth of the modes).
In passing we stress that even for very slow rotators (rotational splitting as low as a few $\mu$Hz) the $m\neq0$ components of non-radial multiplets
are not expected to be symmetrical with respect to the $m=0$ component. We proposed a simple way of treating this assymetry in future analyses of solar-like pulsators.

The p modes of \cible\ were found to have lifetimes ranging from about one day to two days, \textit{i.e.} somewhat shorter than the mode lifetimes in the Sun, but significantly larger than those
of the previously observed \corot\ solar-like targets. This explains why the spectrum of \cible\ is much clearer than for previous \corot\ pulsators, for which the large mode linewidths
made it harder to separate the $\ell=2$ ridge from the $\ell=0$ ridge. The results obtained for \cible\ confirm that the linewidths of the modes in G-type pulsators are
smaller than those of F-type pulsators, making their analysis easier.

The very high quality of the spectrum also enabled us to detect significant peaks outside the identified ridges. These
peaks were found to be compatible with mixed modes, whose presence can be expected in the spectrum of evolved objects
such as \cible. 
The existence of mixed modes in avoided crossing can explain some specific behaviors we observe in the low-frequency 
eigenmodes. In particular we found that the $\ell=1$ ridge is distorted compared to the $\ell=0$ ridge at low frequency.
\cite{deheuvels10} showed that this type of distortion could be associated with
a low-degree mixed mode in avoided crossing. The identification of the
mode $\pi_1$ as an $\ell=1$ mixed mode would be consistent with the observed pattern. 
Other features like the $\ell=2$ mode 
found to overlap the $\ell=0$ mode (around 855 $\mu$Hz) might also result from avoided crossing phenomena.
This needs to be further investigated in the seismic interpretation of this star.
%we observe a distortion in the $\ell=1$ ridge at low frequency which is very similar to the one
%we would expect in the case of an avoided crossing in the vicinity of one of these modes (see \cite{deheuvels10}).
%This needs to be investigated }

%This makes \cible\ a very promising object regarding the study of the structure of solar-like pulsators.

%______________________________________________________________
\begin{acknowledgements}
This work was supported by the Centre National d'Etudes Spatiales (CNES).
We thank M.-A. Dupret for providing estimates of the mode linewidths for \cible, computed with the non-adiabatic pulstion code MAD.
I.~W. Roxburgh and G. Verner thank the UK Science and
Technology Facilities Council for support under grant PP/E001793/1
\end{acknowledgements}

% for the bibliography, at the end
\bibliographystyle{aa.bst} % style aa.bst
\bibliography{biblio} % your references Yourfile.bib

\begin{appendix}

\section{Estimate of the expected amplitudes for the oscillations in \cible \label{app_amp}}
Scaling laws give an estimate of the amplitude of $\ell=0$ modes at the maximum of the signal.
\cite{1995A&A...293...87K} established the following relation between the luminosity amplitude $A$ of the oscillation and the velocity amplitude $v\ind{osc}$:
\begin{equation}
A\equiv \frac{\delta L}{L} \varpropto \frac{v\ind{osc}}{\sqrt{T\ind{eff}}}.
\label{eq_amp}
\end{equation}
We assume that the frequency at the maximum of the signal $\nu\ind{max}$ scales as the acoustic cutoff frequency $\nu\ind{c}$, \textit{i.e.} 
\begin{equation}
\nu\ind{max} \varpropto \frac{g}{\sqrt{T\ind{eff}}} \varpropto \frac{M}{R^2\sqrt{T\ind{eff}}}.
\label{eq_kjeld}
\end{equation}
\cite{samadi07} obtained
\begin{equation}
v\ind{osc} \varpropto \left( \frac{L}{M} \right) ^{0.7} \varpropto \left( \frac{R^2{T\ind{eff}}^4}{M} \right) ^{0.7}.
\label{eq_samadi}
\end{equation}
Inserting Eq. \ref{eq_kjeld} and \ref{eq_samadi} into Eq. \ref{eq_amp} we obtain
\begin{equation}
A=A_{\odot}\left( \frac{T\ind{eff}}{T_{\odot}} \right)^{1.95}\left( \frac{\nu\ind{max}}{\nu_{\rm{max}\odot}} \right)^{-0.7}.
\end{equation}
With the spectroscopic measure of the temperature $T\ind{eff}=6095\pm50$ K and our estimate $\nu\ind{max}=1013\pm3\,\mu$Hz we obtain the expected
amplitude for $\ell=0$ modes at the signal maximum $A_{\ell=0}=6.1\pm0.5$ ppm.

\section{Computation of the posterior probability \label{app_bayes}}
By applying the Bayes theorem and assuming an equiprobability of H$_0$ and H$_1$, \cite{app09} obtained the following expression for the posterior probability
of the H$_0$ hypothesis given the observed data:
\begin{equation}
P(H_0|x)=\left( 1+\frac{P(x|H_1)}{P(x|H_0)} \right) ^{-1}.
\label{eq_bayes}
\end{equation}
The expression of $P(x|H_0)$ for a spectrum binned over $n$ bins is given in \cite{app04}
\begin{equation}
P(x|H_0)=\frac{x^{n-1}e^{-x}}{\gamma(n)},
\label{eq_p0}
\end{equation}
where $\gamma(n)$ is the Gamma function.
To solve Eq. \ref{eq_bayes}, we need to define the alternate hypothesis H$_1$ and
to make prior assumptions on the expected signal. 
\\

Among the four studied peaks, two are assumed to be $\ell=3$ modes around the maximum of the signal ($\pi_2$ and $\pi_3$). 
To obtain an expression for $P(x|H_1)$, we need an estimate of the expected height and linewidth of $\ell=3$ modes. 

An estimate of the expected mode linewidths of \cible\ was computed with the non-adiabatic pulsation code MAD combined with a 1D model reproducing
the effective temperature, the luminosity and the $\log g$ of the object in the same way as done in \cite{app09}.
We obtain around the maximum of the signal a linewidth between
1 $\mu$Hz and 2.5 $\mu$Hz. We adopt a uniform prior for the linewidth, taking into account an uncertainty factor in the theoretical model:
we assume that the maximum linewidth is twice as large as that given from the models ($\Gamma \in$[$0,5$]~$\mu$Hz).

The height of the $\ell=3$ modes can be deduced from the amplitude of $\ell=0$ modes through the relation
\begin{equation}
H_{\ell,m}=r_{\ell,m}(i)H_{\ell=0}=r_{\ell,m}(i)\frac{{A_{\ell=0}}^2}{\pi \Gamma}
\end{equation}
for an angle of inclination $i$. The visibility factors $r_{\ell,m}(i)$ are computed from \cite{2003ApJ...589.1009G}. The expected amplitude of the $\ell=0$
modes around the maximum of the signal is computed in Appendix \ref{app_amp}. We impose a uniform prior for the amplitude of the $\ell=0$ modes 
($A\in$ [$A\ind{min},A\ind{max}$], where $A\ind{min}$ and $A\ind{max}$ are the theoretical values of $A_{\ell=0}$ at $\pm3\sigma$) and
for the angle of inclination ($i\in$ [$0,\pi/2$]).

Based on \cite{app09} we obtain the following expression for the probability of the observed data given the signal
\begin{equation}
P(x|H_1)=\frac{1}{\eta}\int_{A\ind{min}}^{A\ind{max}}\int_0^{\Gamma\ind{max}}\int_{0}^{\pi/2} \frac{\lambda^{\nu}}{\gamma(\nu)}x^{\nu-1}e^{-\lambda x}\rm{d}A'\rm{d}\Gamma '\rm{d}i',
\label{eq_p1}
\end{equation}
where $\eta=\Gamma\ind{max}(A\ind{max}-A\ind{min})\pi/2$ is the normalization factor. $\lambda$ and $\nu$ are given in \cite{app04} and are functions of the
mode height $H$ and linewidth $\Gamma$.

Inserting Eq. \ref{eq_p0} and \ref{eq_p1} into Eq. \ref{eq_bayes}, we obtain for $\pi_2$ and $\pi_3$ a posterior probability of 1.4\% and 1.1\%, respectively.
\\

Two peaks are assumed to be the signature of mixed modes ($\pi_1$ and $\pi_4$). For this type of modes, a part of the energy is located in the g-mode cavity and the inertia is larger.
We therefore expect a smaller linewidth for these modes. But we can have no clue of their degree $\ell$.
Because both their amplitudes and their linewidths greatly depend on $\ell$, we use very conservative priors: uniform priors on the amplitude of the mode
($A\in$ [$0,A\ind{max}$] ppm) and on the linewidth ($\Gamma \in$[$0,5$] $\mu$Hz).

We obtain posterior probabilities of 0.001\% and 8.7\% for $\pi_1$ and $\pi_4$, respectively.

\end{appendix}

\end{document}